\documentclass[10pt,twocolumn]{IEEEtran}

\usepackage{amsmath,graphics,amssymb,epsfig,color,cite,comment}

\usepackage{amsthm}
\theoremstyle{plain}
\newtheoremstyle{mystyle}
  {0mm}
  {0mm}
  {}
  {4mm}
  {\bfseries}
  {:}
  { }
  {\thmname{#1}\thmnumber{ #2}\thmnote{ (#3)}}
\theoremstyle{mystyle}

\usepackage{array}
\usepackage{multirow}
\usepackage{enumerate}
\usepackage{bm}
\usepackage{float}
\usepackage{makecell}
\usepackage[right=0.625in, left=0.625in, top=0.7in, bottom=1.1in]{geometry}

\usepackage{algorithm2e}
\usepackage{algpseudocode}
\usepackage{algorithmicx}
\usepackage{algcompatible}
\usepackage{setspace}

\usepackage[caption=false,font=footnotesize]{subfig}
\usepackage{bbm}
\usepackage{bigints}
\usepackage{adjustbox}

\floatname{algorithm}{Algorithm}

\usepackage{enumitem}

\newcommand{\argmin}{\operatornamewithlimits{argmin}}
\makeatletter
\newcommand{\vast}{\bBigg@{4.5}}
\newcommand{\Vast}{\bBigg@{7.5}}
\makeatother

\begin{document}
\title{\fontsize{23}{28}\selectfont Rate-Adaptive Semantic Communication via Multi-Stage Vector Quantization
}

\author{Jinsung Park, Junyong Shin, Yongjeong Oh, Jihun Park, and Yo-Seb Jeon
	    \thanks{Jinsung Park, Junyong Shin, Yongjeong Oh, Jihun Park, and Yo-Seb Jeon are with the Department of Electrical Engineering, POSTECH, Pohang, Gyeongbuk 37673, Republic of Korea (e-mail: jinsung@postech.ac.kr; sjyong@postech.ac.kr; yongjeongoh@postech.ac.kr; jihun.park@postech.ac.kr; yoseb.jeon@postech.ac.kr).}
	}\vspace{-2mm}	
	
	\maketitle
	\vspace{-12mm}

\begin{abstract} 
This paper proposes a novel framework for rate-adaptive semantic communication based on multi-stage vector quantization (VQ), termed \textit{MSVQ-SC}. Unlike conventional single-stage VQ approaches, which require exponentially larger codebooks to achieve higher fidelity, the proposed framework decomposes the quantization process into multiple stages and dynamically activates both stages and individual VQ modules. This design enables fine-grained rate adaptation under varying bit constraints while mitigating computational complexity and the codebook collapse problem. To optimize performance, we formulate a module selection problem that minimizes task loss subject to a rate constraint and solve it using an incremental allocation algorithm. Furthermore, we extend the framework by incorporating entropy coding to exploit non-uniform codeword distributions, further reducing communication overhead. Simulation results on the CIFAR-10 dataset demonstrate that the proposed framework outperforms existing digital semantic communication methods, achieving superior semantic fidelity with lower complexity while providing flexible and fine-grained rate control.
\end{abstract}


\section{Introduction}\label{Sec:Intro}


Semantic communication (SC) has recently emerged as a paradigm that transmits the semantic meaning of information rather than its raw data representation. Unlike conventional systems that treat all bits uniformly, SC prioritizes task-relevant information, thereby reducing redundancy and enabling efficient utilization of communication resources \cite{SC_1, SC_2, SC_3, SC_4}. Such efficiency is particularly advantageous in data-intensive scenarios, including autonomous driving, industrial IoT, and real-time multimedia services. By emphasizing semantic relevance over bit-level fidelity, SC introduces a fundamentally new design principle that directly aligns communication with task performance.

Early research on SC has primarily focused on analog implementations. In this paradigm, semantic features are directly mapped into continuous analog symbols and transmitted without conventional channel coding or digital modulation. For example, a Transformer-based SC method was developed in \cite{Anlg_1} for text transmission, while a CNN-based joint source-channel coding (JSCC) scheme was introduced in \cite{Anlg_2} for image transmission. These pioneering works highlighted the potential of analog SC for end-to-end semantic transmission.
However, these early methods focus on supporting a fixed communication rate, which limits flexibility and adaptability when communication environments change. 

In practice, wireless channels are highly dynamic: signal quality can fluctuate due to fading, interference, or user mobility, and the available bandwidth is frequently constrained by competing services and varying network conditions. Under such variability, relying on a fixed rate setting can lead either to inefficient resource utilization, when the rate is higher than necessary, or to noticeable degradation in reconstruction quality, when the rate is insufficient. This challenge highlights the importance of adaptively adjusting the transmission rate to match both the current channel environment and the specific task requirements.

\subsection{Rate-Adaptive Semantic Communication}
To address this issue, rate-adaptive SC has been investigated, with the goal of dynamically adjusting transmission rates (or related parameters) according to channel conditions and task requirements. For example, the authors in \cite{adapt_NN1,adapt_NN2} developed rate-adaptive SC frameworks that control the dimension of the semantic feature. In these approaches, the transmission rate is determined prior to feature extraction based on channel state information (CSI), enabling the encoder–decoder pair to accommodate multiple rates and thus adapt flexibly to varying channel conditions. Although these methods support dynamic rate control, they require separate encoder–decoder pairs for each supported rate, with each pair trained independently. As a result, achieving finer rate granularity necessitates additional branches, which in turn increases model complexity and memory overhead.

Another representative approach for rate-adaptive SC is to employ masking \cite{adapt_Mask_1, adapt_Mask_2, adapt_Mask_3, adapt_Mask_4}. In this approach, masking is performed by a neural network that leverages statistical CSI to select which semantic features are transmitted. In \cite{adapt_Mask_1, adapt_Mask_2, adapt_Mask_3}, network modules employing the Gumbel–Softmax technique were developed to dynamically regulate the transmission rate in response to channel variations. Although these studies provide rate-adaptive mechanisms, their training procedures are restricted to a finite set of predefined rates, thereby limiting the granularity and flexibility of rate control. Unlike these works, an importance-driven approach was proposed in \cite{adapt_Mask_4}, where feature scores are computed and low-importance features are masked to achieve rate adaptation. However, estimating feature importance requires gradient- or error-based evaluation, which incurs considerable computational overhead, especially for high-dimensional features.

The common limitation of all the aforementioned rate-adaptive methods is that they rely on analog transmission of semantic features. Such analog SC approaches face inherent drawbacks: direct analog mapping makes the system vulnerable to channel noise, interference, and circuit imperfections, while also being incompatible with modern wireless standards that rely on digital communication. Furthermore, error correction and retransmission mechanisms, essential for reliable communication, are difficult to integrate into purely analog pipelines.

\subsection{Digital Semantic Communication}
Recent research has increasingly focused on digital SC to overcome the inherent limitations of analog SC. Unlike analog approaches, digital SC encodes semantic features into discrete representations, which can be further protected by channel coding and error correction. This design enhances robustness against noise and interference while ensuring compatibility with modern wireless standards built on digital infrastructures. To realize these benefits, several digital SC frameworks have been proposed that embed semantic processing within conventional digital pipelines.

A widely adopted strategy in digital SC is to employ a quantization module that maps continuous semantic features into discrete indices. These indices are then encoded into bit sequences and transmitted using standard digital modulation and channel coding techniques. Two representative quantization strategies have been explored. Scalar quantization (SQ) discretizes each feature dimension independently, offering simplicity and low computational complexity, which makes it attractive for lightweight devices and latency-sensitive applications \cite{Jihun:ICTExpress, BSC_Goldsmith}. In contrast, vector quantization (VQ) jointly maps groups of feature values to codewords in a codebook. Since VQ captures cross-dimensional correlations and provides more compact representations, it generally achieves superior performance compared to SQ in digital SC systems \cite{Shin:ESC-MVQ}. In addition, the VQ approach can be extended to support multiple transmission rates, by adopting nested codebooks, as studied in \cite{nested}.
However, most VQ-based SC frameworks rely on a single-stage quantization mechanism. In such setups, achieving high semantic fidelity requires significantly larger codebooks, which increases computational complexity due to expensive nearest-neighbor searches. Moreover, excessively large codebooks often suffer from sparse codeword utilization, leading to inefficient representation and the well-known codebook collapse problem \cite{codebook_collapse}.

Multi-stage vector quantization (MSVQ) has been proposed as a promising solution to mitigate the limitations of the single-stage VQ approach. MSVQ decomposes the quantization process into multiple stages, where each stage quantizes the residual error of the previous one. As the number of stages increases, the overall quantization error is progressively reduced. This design reduces the need for very large codebooks at each stage, alleviates codebook collapse, and achieves a favorable balance between representational fidelity and computational complexity \cite{MSVQ0, MSVQ1, MSVQ2, MSVQ3}. 
Notably, MSVQ inherently supports multiple transmission rates by selectively activating different numbers of stages, making it a natural candidate for rate-adaptive digital SC.
Despite this potential, no prior work has systematically investigated rate-adaptive SC based on MSVQ. In particular, when the semantic feature dimension is large, which is a common scenario in SC applications, the simple selection of the number of active stages is insufficient to provide fine-grained rate control. Therefore, achieving flexible and fine-grained rate adaptation in MSVQ-based SC still remains an open and important problem.  

\subsection{Our Contributions}
To address this research gap, we propose a novel framework for rate-adaptive SC based on MSVQ, referred to as \textit{MSVQ-SC}. Unlike existing approaches, the proposed framework regulates the bit overhead not only at the stage level but also at the individual VQ module level, thereby enabling fine-grained rate adaptation. Furthermore, we formulate and solve an optimization problem that determines the selection of VQ modules across stages by explicitly accounting for their impact on task performance. The main contributions of this paper are summarized as follows:
\begin{itemize}

    \item 
    
    We propose a novel rate-adaptive communication strategy for the MSVQ-based SC. Unlike prior studies that restrict rate adaptation to coarse stage-level control, our approach enables fine-grained rate control by dynamically adjusting the activation of both stages and individual VQ modules. This mechanism significantly improves flexibility in adapting to diverse channel conditions and enhances the efficiency of semantic information transmission. To the best of our knowledge, this is the first work to investigate module-level rate control in MSVQ-based SC.

    \item 
    We formulate an optimization problem to determine the optimal selection of activated VQ modules that minimizes task loss under a given rate constraint. To solve this problem, we adopt an incremental allocation algorithm in \cite{Fox} and specify the conditions under which its optimality is guaranteed. 
    It should be noted that the problem of optimizing module selection in MSVQ-based SC has never been explored in the literature.

    \item 
    We further extend the proposed MSVQ-SC framework by incorporating entropy coding into each VQ module. In this extension, module selection is refined by considering the average bit overhead derived from the codeword distribution of each module. This integration reduces communication overhead while maintaining adaptive rate control.

    \item 
    Through simulations on the CIFAR-10 dataset, we demonstrate that the proposed MSVQ-SC framework outperforms existing digital SC methods while offering fine-grained rate control across a wide range of operating points. Moreover, we show that the incorporation of entropy coding effectively enhances the transmission efficiency of the proposed framework. 

\end{itemize}




\section{System Model and Preliminary}\label{Sec:System Model and Preliminary}
In this section, we present a digital SC system considered in our work and introduce an existing single-stage VQ architecture adopted in the literature.

\subsection{Digital SC Systems}\label{Sec:System}

We consider a digital SC system in which a receiver intends to perform an inference task using the received representation of the source data ${{\bf x}}\in \mathbb{R}^{K}$ sent from a transmitter. The considered system is illustrated in Fig. 1.
\begin{figure}[t]
    \centering
    {\epsfig{file=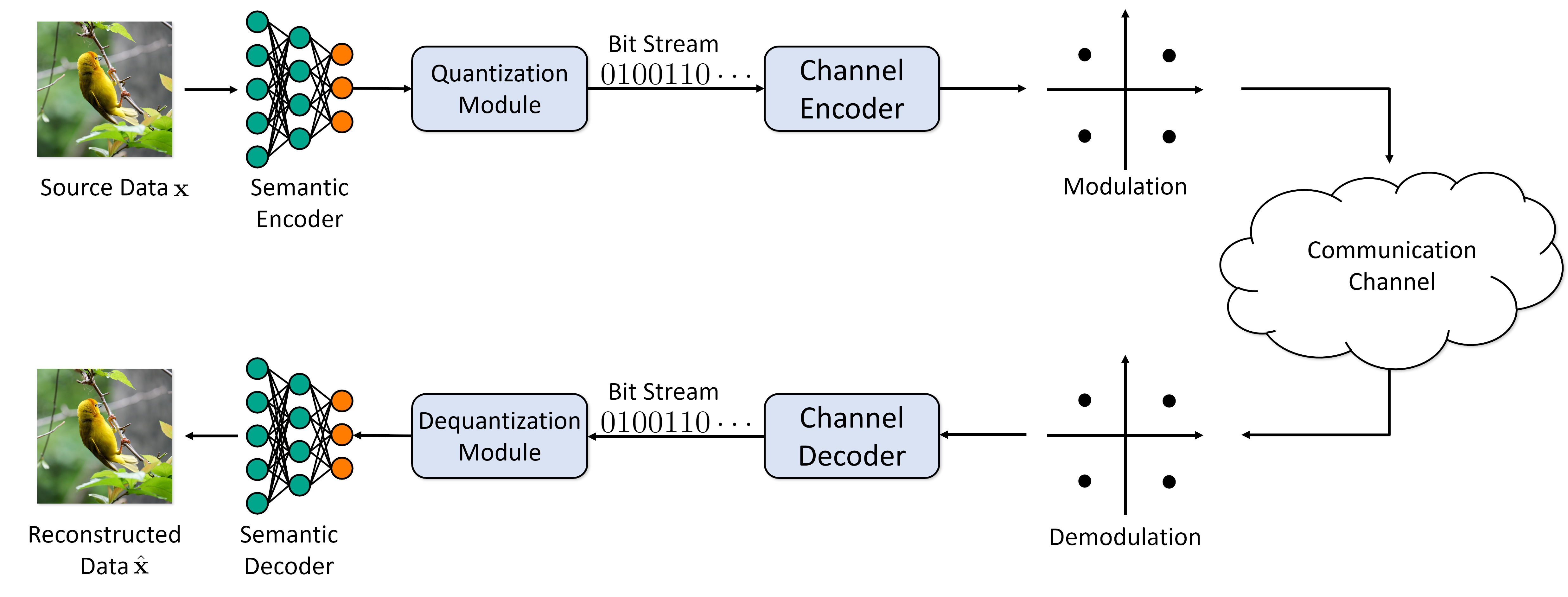,width=9cm}}
    \caption{An illustration of the digital SC system considered in this work, where a receiver intends to perform an inference task  using the received representation of the source data ${{\bf x}}\in \mathbb{R}^{K}$ sent from a transmitter.}\vspace{-3mm}
    
    \label{fig:MSVQVAE_v4}
\end{figure}

At the transmitter, a semantic encoder $f_{\bm \theta}({\cdot})$, parameterized by ${\bm \theta}$, is employed to extract high-level semantic features from the input data. 
As a result, a lower-dimensional latent representation ${\bf z}\in {\mathbb R}^{M}$ is obtained as follows:
\begin{align}\label{eq: x-z}
{\mathbf{z}} = f_{\bm \theta}({{\bf x}})\in \mathbb{R}^{M},
\end{align}
where $M$ is the dimension of the semantic feature vector. The output $\bf z$ serves as a compressed semantic representation that captures task-relevant features of the input data. To transmit the semantic feature vector ${\bf z}$ using digital communication, a quantization module $Q:{\mathbb R}^{M} \rightarrow \{0,1\}^{B_{\rm tot}}$ is employed which transforms  ${\bf z}$ into a finite-bit sequence with length ${B}_{\rm tot}$. 
The resulting bit sequence is transmitted following a typical digital modulation process. For example, the bit sequence can be modulated into a symbol sequence using QAM signaling and then transmitted with appropriate power allocation. 

At the receiver, a transmitted bit sequence is reconstructed via a typical digital demodulation process which may include channel estimation, channel equalization, symbol demodulation, and channel decoding. 
After this process, the reconstructed bit sequence is converted into the semantic feature vector, denoted by $\hat{\bf z}$. Note that due to the quantization process applied at the transmitter, the reconstructed vector $\hat{\bf z}$ may differ from the original feature semantic vector ${\bf z}$. 
The reconstructed latent vector $\hat{\bf z}$ is then passed through a semantic decoder $g_{{\bm \phi}}(\cdot)$, parameterized by $\bm \phi$, which aims at reconstructing inference-related data from the semantic feature vector.
For example, when considering an image reconstruction task, the output of the semantic decoder is given by 
\begin{align}\label{eq: z-x}
    {\hat{{\bf x}}} = g_{{\bm \phi}}(\hat{\mathbf{z}})\in \mathbb{R}^{K},
\end{align}
where ${\hat{{\bf x}}}$ denotes the reconstructed source data. 

\subsection{Existing Single-Stage VQ Architecture}\label{Sec:Single_VQ}
One of the main factors that govern the performance of the receiver's task is the design of a quantization module. This quantization module needs to properly capture the distribution of the semantic feature vector. To this end,  a single-stage VQ architecture has been adopted in the literature \cite{shape-gain,ECVQ, ICTC, Wi-Fi}, which utilizes a learnable VQ module that captures the distribution of the semantic features in a data-driven manner.

In this architecture, the semantic feature vector $\bf z$ is first partitioned into $N$ sub-vectors, each with dimension $D$, in order to avoid high computational complexity required when quantizing a high-dimensional semantic feature vector. Details of our strategy to divide ${\bf z}$ into $N$ sub-vectors will be discussed in Sec.~III-B.
Let ${\bf z}_{i}\in {\mathbb R}^{D}$ be the $i$-th feature sub-vector, where $i \in\{1,\ldots,N\}$ is the index corresponding to each sub-vector. 
Note that the feature dimension $M$ holds $M=N\times D$.
Each sub-vector is quantized to $\mathbf{z}_{q,i}$ according to the minimum Euclidean by selecting the nearest codeword $\mathbf{c}_k$ in $\mathcal{C}_{\rm single}$ based on the Euclidean distance:
\begin{align}\label{eq: quantize}
    \hat{\bf z}_{i} =\argmin_{\mathbf{c}_k \in \mathcal{C}_{\rm single}}\Vert\mathbf{z}_i - \mathbf{c}_k \Vert^2.
\end{align}
By concatenating the quantized sub-vectors, the quantized semantic feature is obtained as
\begin{align}
    \hat{\bf z} =\big[\hat{\bf z}_{1}^{\sf T},\cdots,\hat{\bf z}_{N}^{\sf T}\big]^{\sf T}.
\end{align}

Although the single-stage VQ architecture discussed above supports a learnable VQ codebook framework in \cite{VQ0} to capture the distribution of the semantic feature vector, it faces significant limitations for rate-adaptive SC. A key drawback is that the reduction of the quantization error is achieved solely by increasing the size of the VQ codebook, $|\mathcal{C}_{\rm single}|$, which in turn leads to a substantial increase in the computational complexity of the quantization process. In addition, as the codebook size grows, the utilization of codeword vectors becomes saturated---a phenomenon known as the codebook collapse problem \cite{codebook_collapse}. Consequently, this architecture exhibits limited scalability in supporting high-rate communication scenarios.

\section{Proposed Multi-Stage VQ-based Semantic Communications (MSVQ-SC)}\label{Sec:MSVQ-SC}


In this section, we propose a digital SC framework based on MSVQ, referred to as MSVQ-SC. The proposed framework provides learnable and efficient quantization of the semantic feature vector while enabling flexible and fine-grained rate control.

\begin{figure}[t]
    \centering
    {\epsfig{file=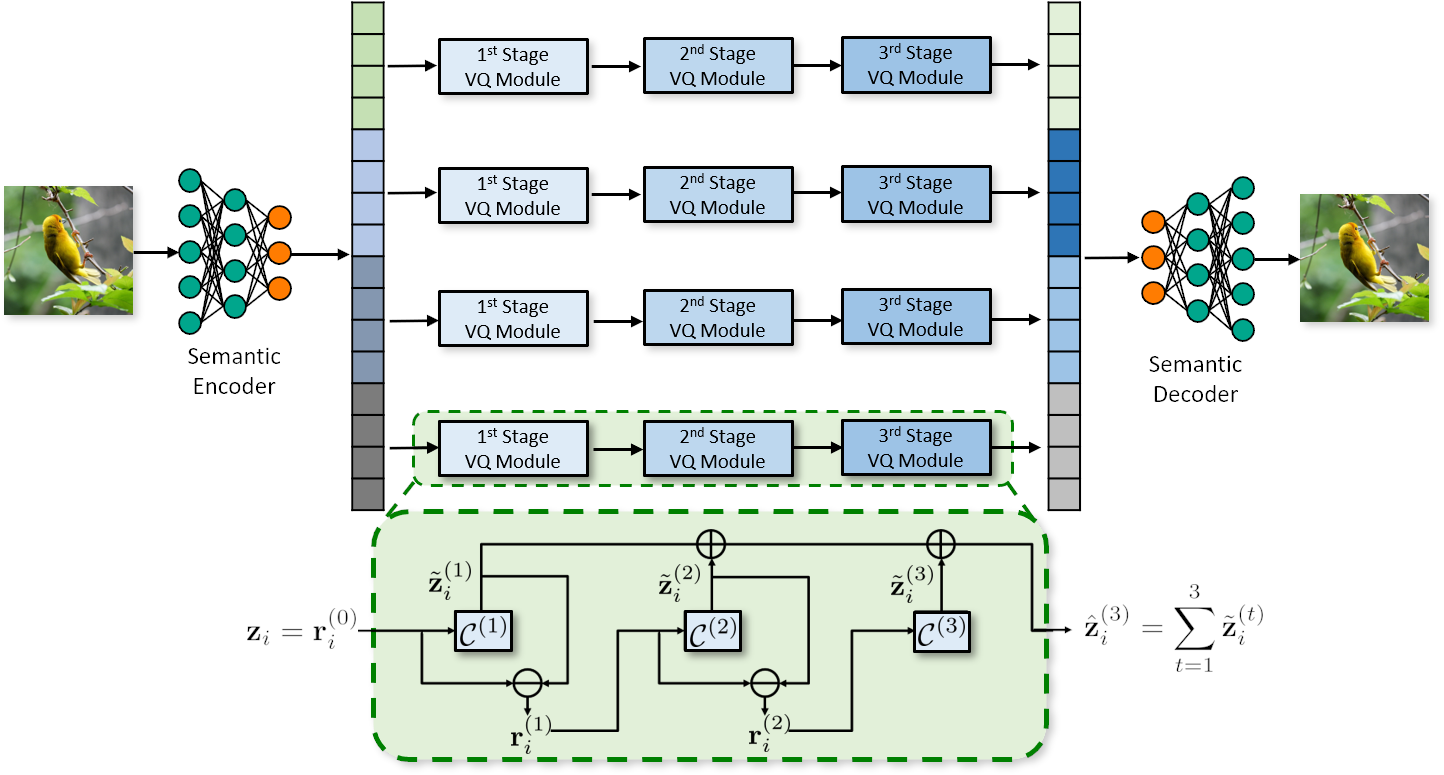,width=9cm}}
    \caption{Illustration of multi-stage VQVAE.}\vspace{-3mm}
    
    \label{fig:MSVQVAE_v4}
\end{figure}

\subsection{Digital SC with MSVQ}
In the proposed framework, we employ an MSVQ architecture as the quantization module for the semantic feature vector. The MSVQ architecture enables a gradual reduction in quantization error without significantly increasing the VQ codebook size, thereby effectively addressing the limitations of the conventional single-stage VQ architecture introduced in Sec.~\ref{Sec:Single_VQ}.

In this architecture, multiple VQ modules connected in cascade are used to quantize each feature sub-vector ${\bf z}_i$, as illustrated in Fig. \ref{fig:MSVQVAE_v4}.
Let $T_{\rm max}>1$ denote the total number of VQ modules available for quantizing each sub-vector. The VQ module in each stage $t \in \{1,\ldots,T_{\rm max}\}$ quantizes a \emph{residual} vector produced at the previous stage  $(t-1)$. For the $i$-th sub-vector ${\bf z}_i$ at stage $t$, the residual vector ${\bf r}_i^{(t)} \in \mathbb{R}^D$ is defined as the difference between the previous residual vector ${\bf r}_{i}^{(t-1)}$ and the quantization output ${\tilde{\bf z}}_{i}^{(t)}$ as follows:
\begin{align}\label{eq: VQ_input}
    {\bf r}_i^{(t)} = {\bf r}_i^{(t-1)} - \tilde{\bf z}_i^{(t)}.
\end{align}
The initial residual vector is defined as ${\bf r}_{i}^{(0)}={\bf z}_{i}$. 

At stage $t$, the residual vector ${\bf r}_{i}^{(t-1)}$ is quantized by searching for the closest codeword in the Euclidean sense:
\begin{align}\label{eq:VQ_quantize}
    \tilde{\bf z}_{i}^{(t)}=\argmin_{\mathbf{c}_k \in \mathcal{C}_i^{(t)}}\Vert\mathbf{r}_i^{(t-1)} - \mathbf{c}_k \Vert^2.
\end{align}
where $\mathcal{C}_i^{(t)}$ is the VQ codebook for the $i$-th sub-vector in stage $t$, which consists of $K_i^{(t)}\triangleq 2^{B_i^{(t)}}$ codeword vectors, and $B_i^{(t)} \in \mathbb{N}$ the quantization bits allocated to the $i$-th sub-vector in stage $t$.
Then, the quantized representation of the $i$-th sub-vector is obtained by summing the outputs of the VQ modules up to stage $t$:
\begin{align}\label{eq: Output}
    \hat{\bf z}_{i}^{(t)} = \sum_{t^\prime=1}^{t}  \tilde{\bf z}_i^{(t^\prime)}. 
\end{align}
Accordingly, the residual error between the original sub-vector and its quantized representation at stage $t$ is given by
\begin{align}\label{eq: Total_error}
    {\bf r}_i^{(t)} = {\bf z}_i - \hat{\bf z}_{i}^{(t)} = {\bf z}_i -\sum_{t^\prime=1}^{t}  \tilde{\bf z}_i^{(t^\prime)}.
\end{align}

The expression in \eqref{eq: Total_error} implies that the quantization error decreases progressively as the number of the stages increases. Consequently, the more stages that are used, the closer the final quantized representation approximates the original semantic feature vector in the embedding space. This observation underscores the importance of effectively training the quantization modules at each stage, ensuring that each codebook learns to capture and refine the remaining information in the residual vectors. High-quality VQ codebooks across stages are therefore essential for minimizing distortion and achieving high-fidelity reconstructions within the MSVQ framework.

\subsection{Initial Training for Sub-Vector Construction}
In our approach, proper construction of the $N$ sub-vectors from the semantic feature vector is crucial for enabling effective quantization. It is well known that the quantization error of a given signal is closely related to its variance  \cite{std}. Furthermore, \cite{std} has shown that neural network outputs with higher variances tend to correspond to more important features during training. Motivated by these insights, we group entries with similar variances into the same sub-vector. This allows sub-vectors containing high-variance entries to be properly prioritized in subsequent quantization and transmission processes.

To achieve this, we first quantify the intrinsic variance of each entry of the semantic feature vector $\bf z$. 
However, since in our framework the codebooks are jointly trained with the semantic encoder and decoder, the variance of each entry inherently depends on the codebook design. To avoid this entanglement, we introduce an initial training process in which the semantic encoder-decoder pair is trained \emph{without} the quantization module. In this process, the semantic encoder's output directly serves as the input to the semantic decoder. Let ${\bm \theta}_{\rm ini}$ and ${\bm \phi}_{\rm ini}$ denote the encoder and decoder parameters during the initial training.  The training loss used in this stage is given by
\begin{align}\label{eq: Initial Loss}
    \mathcal{L}_{\rm ini}({\bm \theta}_{\rm ini},{\bm \phi}_{\rm ini}) &=  \left\Vert {\bf x}- g_{{\bm \phi}_{\rm ini}}(f_{{\bm \theta}_{\rm ini}}({{\bf x}}))\right\Vert^2,
\end{align}
After the initial training process, we compute the variance of each entry of the resulting semantic feature vector $\bf z$ using the training dataset.
The variance of the $m$-th entry of ${\bf z}$ is expressed as
\begin{align}\label{position variance}
    \sigma_m^2 = \frac{1}{|\mathcal{D}|}\sum_{{\bf x}^\prime \in\mathcal{D}} \|\big(f_{{\bm \theta}_{\rm ini}}({{\bf x}^\prime})\big)_m - \mu_m\|^2,
\end{align}
where $\mathcal{D}$ represents the training dataset, and 
\begin{align}
    \mu_m = \frac{1}{|\mathcal{D}|}\sum_{{\bf x}^\prime \in\mathcal{D}} \big(f_{{\bm \theta}_{\rm ini}}({{\bf x}^\prime})\big)_m.
\end{align}

Based on these computed variances, we divide ${\bf z}$ into $N$ sub-vectors so that each sub-vector comprises semantic feature entries with similar variances. Specifically, the $i$-th sub-vector ${\bf z}_i$ is constructed as
\begin{align}\label{eq:sub-vector_construct}
    {\bf z}_i= \big[ ({\bf z})_{m_{(i-1)D+1}^{\prime}},\cdots,  ({\bf z})_{m_{iD}^{\prime}}\big]^{\sf T},~\forall i \in [N]
\end{align}
where $m^{\prime}_i$ denotes the index of the $i$-th largest variance among $\sigma_1^2, \sigma_2^2, \ldots, \sigma_M^2$.

{\bf Remark 1 (Quantization Bit Allocation):}
Our sub-vector construction also provides intuitive guidelines for allocating different quantization bits across sub-vectors and stages. In general, entries of the semantic feature vector with higher variance require higher quantization bits to achieve comparable quantization accuracy. Motivated by this observation, we allocate higher quantization bits to sub-vectors composed of higher-variance entries, following the constraint:
\begin{align}
    B_1^{(t)} \geq B_2^{(t)} \geq \ldots \geq B_N^{(t)},~\forall t\in [T_{\rm max}].
\end{align}
Similarly, since the quantization inputs in earlier stages naturally exhibit larger variances compared to those in later stages, we allocate higher quantization bits to modules in the earlier stages, following the constraint:
\begin{align}
    B_i^{(1)} \geq B_i^{(2)} \geq \ldots \geq B_i^{(T_{\rm max})},~\forall i\in [N].
\end{align}
Guided by these constraints, we determine the set of quantization bits $\{B_i^{(t)}\}_{\forall i,t}$ which will be explicitly specified in Sec. VI.

\subsection{Joint Training of Semantic Encoder, VQ Modules, and Semantic Decoder}\label{Joint_Training}
After the sub-vector construction process, MSVQ modules are jointly trained with the semantic encoder and decoder, in order to enable efficient finite-bit representation while guaranteeing the task performance. For the joint training process, we consider two strategies, one-shot training and sequential training. 
The one-shot strategy optimizes all these modules simultaneously, while the sequential strategy adopts a progressive approach by training the VQ modules for stages incrementally up to $T_{\rm max}$.
In both strategies, we initialize the encoder and decoder using the parameters ${\bm \theta}_{\rm ini}$ and ${\bm \phi}_{\rm ini}$, respectively, obtained from the initial training stage. 
In what follows, for notational simplicity, we define $\bar{\mathcal{C}}^{(t)}$ as a set of the VQ codebooks used in the MSVQ operation up to stage $t$, i.e., 
\begin{align}
    \bar{\mathcal{C}}^{(t)} = \{\mathcal{C}_i^{(t^\prime)}\}_{i\in[N],t^\prime \leq t},~\forall t\in [T_{\rm max}].
\end{align}
We also define $Q_{\bar{\mathcal{C}}^{(t)}}^{(t)}({\bf z})=\hat{\bf z}^{(t)}$ as the output of the VQ modules up to stage $t$ for the input feature vector ${\bf z}$. We assume that the output $\hat{\bf z}^{(t)}$ is obtained by rearranging $N$ quantized sub-vectors $\{\hat{\bf z}_{i}^{(t)}\}_{i\in[N]}$.


\begin{itemize}

    \item {\bf One-Shot Training:} 
    In one-shot training, the encoder, decoder, and all quantization modules in an architecture composed of $T_{\rm max}$ stages are jointly trained using the full configuration. 
    Although the model supports multiple configurations corresponding to different rates, only the complete configuration, consisting of the encoder, all $T_{\rm max}$ stages, and the decoder, is considered during training. All modules in this path are optimized in a single end-to-end process. 
    To support this scheme, motivated by the loss function in \cite{VQ0}, we define a stage-wise loss function for a configuration utilizing up to stage $t$ as
        \begin{align}\label{eq: VQ-VAE Loss}
        \mathcal{L}_{t}({\bm \theta},{\bm \phi},\bar{\mathcal{C}}^{(t)}) = & \left\Vert {\bf x}- g_{{\bm \phi}}(Q_{\bar{\mathcal{C}}^{(t)}}^{(t)}(f_{{\bm \theta}}({{\bf x}})))\right\Vert^2\nonumber\\
        &+\left\Vert \text{sg}(f_{{\bm \theta}}({{\bf x}}))-Q_{\bar{\mathcal C}^{(t)}}^{(t)}(f_{{\bm \theta}}({{\bf x}}))\right\Vert^2 \nonumber\\
        &+ \beta \left\Vert f_{{\bm \theta}}({{\bf x}})-\text{sg}(Q_{\bar{\mathcal C}^{(t)}}^{(t)}(f_{{\bm \theta}}({{\bf x}})))\right\Vert^2,
        \end{align}
        where ${\rm sg}(\cdot)$ is a stop-gradient operation, and $\beta$ is a hyperparameter.
        In \eqref{eq: VQ-VAE Loss}, the first term corresponds to the reconstruction loss, while the last two terms correspond to the embedding loss. Since the training at stage $t$ requires fine-tuning the previously learned modules up to stage $(t-1)$, the loss of the one-shot training strategy is given by \cite{MSVQ3}
        \begin{align}\label{eq: one-shot loss}
            \mathcal{L}_{\rm {joint}}=\sum_{t=1}^{T_{\rm max}}w^{(t)}\mathcal{L}_{t}({\bm \theta},{\bm \phi},{\bar{\mathcal C}}^{(t)}).
        \end{align}

    \item {\bf Sequential Training:} Sequential training also uses joint optimization, but in a progressive manner. Specifically, when training the VQ module at stage $t$, we optimize it together with all previously trained modules up to stage $(t-1)$. This allows earlier modules to be fine-tuned for better synergy with the newly added VQ module. During each stage’s training, the encoder and decoder are also optimized jointly. 
    The loss function used for each stage follows the same form as in \eqref{eq: VQ-VAE Loss}. The total loss of the sequential training strategy for a configuration up to stage $t$ is given by \cite{MSVQ3}
    \begin{align}\label{eq: Total Loss}
        \mathcal{L}_{\rm {joint}}^{(t)}=\sum_{j=1}^{t}w^{(j)}\mathcal{L}_{t}({\bm \theta}^{(t)},{\bm \phi}^{(t)},{\bar{\mathcal C}}^{(t)}).
    \end{align}
    Using the above loss, we progressively train the model from the first stage (i.e., $t=1$) up to the final stage (i.e., $t=T_{\rm max}$).

\end{itemize}

The overall training procedure of the proposed MSVQ-SC framework, including initial training, sub-vector construction, and joint training, is summarized in Algorithm~\ref{alg: MSVQ Training}.

{\small \RestyleAlgo{ruled}
    \SetKwComment{Comment}{/* }{ */}
    \begin{algorithm}[h]\label{alg: MSVQ Training}
    \caption{Training Procedure of the Proposed MSVQ-SC Framework}
    \setstretch{1.2}
    \textbf{Initial Training \& Sub-Vector Construction:}\\
    Detach all MSVQ modules\;
    Randomly initialize ${\bm \theta}_{\rm ini}$ and ${\bm \phi}_{\rm ini}$\;
    \For{${\rm epoch} \in  \{0,1,\ldots\}$}{
    Randomly split $D$ into $P$ batches $\{D_p\}_{p=1}^P$;\\
    \For{$p=1,\ldots,P$}{
    Compute $\mathcal{L}_{\rm ini}$ on $D_p$ using \eqref{eq: Initial Loss}\;
    Update ${\bm \theta}_{\rm ini}$ and ${\bm \phi}_{\rm ini}$\;}}  
    Compute $\sigma_m^2$ from \eqref{position variance}, $\forall m$; \\
    Construct sub-vector ${\bf z}_i$ from \eqref{eq:sub-vector_construct}, $\forall i$; \\
    \textbf{Joint Training:}\\
    Attach all MSVQ modules\;
    Set ${\bm \theta}={\bm \theta}_{\rm ini}$ and ${\bm \phi}={\bm \phi}_{\rm ini}$\;
    \If{mode = \textsc{One-shot training}}{
            \For{${\rm epoch} \in \{0,1,\ldots\}$}{
            Randomly split $D$ into $P$ batches $\{D_p\}_{p=1}^P$;\\
            \For{$p=1,\ldots,P$}{
            Compute $\mathcal{L}_{\rm joint}$ using \eqref{eq: one-shot loss}\;
            Update ${\bm \theta}$, ${\bm \phi}$, and $\{\mathcal{C}_i^{(t^\prime)}\}_{i\in[N],t^\prime \in[T_{\rm max}]}$\;}
            }}
    \ElseIf{mode = \textsc{sequential training}}{
            \For{${t} \in \{0,1,\ldots, T_{\rm max}\}$}{
            \For{${\rm epoch} \in\{ 0,1,\ldots\}$}{
            Randomly split $D$ into $P$ batches $\{D_p\}_{p=1}^P$;\\
            \For{$p=1,\ldots,P$}{
            Compute $\mathcal{L}_{\rm joint}^{(t)}$ using \eqref{eq: Total Loss}\;
            Update ${\bm \theta}$, ${\bm \phi}$, and $\{{\mathcal C}_i^{(t')}\}_{i\in[N],t'\in [t]}$\;}
            }
            }
            }


    \end{algorithm}}

\section{Rate-Adaptive Communication Strategy for the Proposed MSVQ-SC Framework}\label{Stage Selection}

In this section, we propose a rate-adaptive communication strategy that enables flexible and fine-grained rate control within the proposed MSVQ-SC framework, allowing it to dynamically adapt to varying rate constraints. To this end, we consider a finite-capacity communication scenario\footnote{The finite-capacity scenario will be considered only for the optimization of the proposed MSVQ-SC framework and will be relaxed when evaluating its performance, as will be shown in Sec.~\ref{Sec:Simulation Results}.} in which $B_{\rm cap} \in \mathbb{N}$ bits can be reliably transmitted using a capacity-achieving channel code in conjunction with appropriate modulation and power allocation schemes. Note that the number of reliably transmitted bits, \( B_{\rm cap} \), is determined by various factors, including transmit power, channel fading, interference, and noise. With this consideration, our focus is on adjusting the length of the transmitted bit sequence, \( B_{\rm tot} \), to satisfy the constraint imposed by \( B_{\rm cap} \). This abstraction also ensures that the applicability of the proposed framework is not restricted to a particular communication process; rather, it can be seamlessly integrated into various systems, such as uncoded or coded multiple-input multiple-output (MIMO) and orthogonal frequency-division multiplexing (OFDM) systems.

\subsection{Main Idea: Stage Selection}

\begin{figure*}[t]
    \centering
    {\epsfig{file=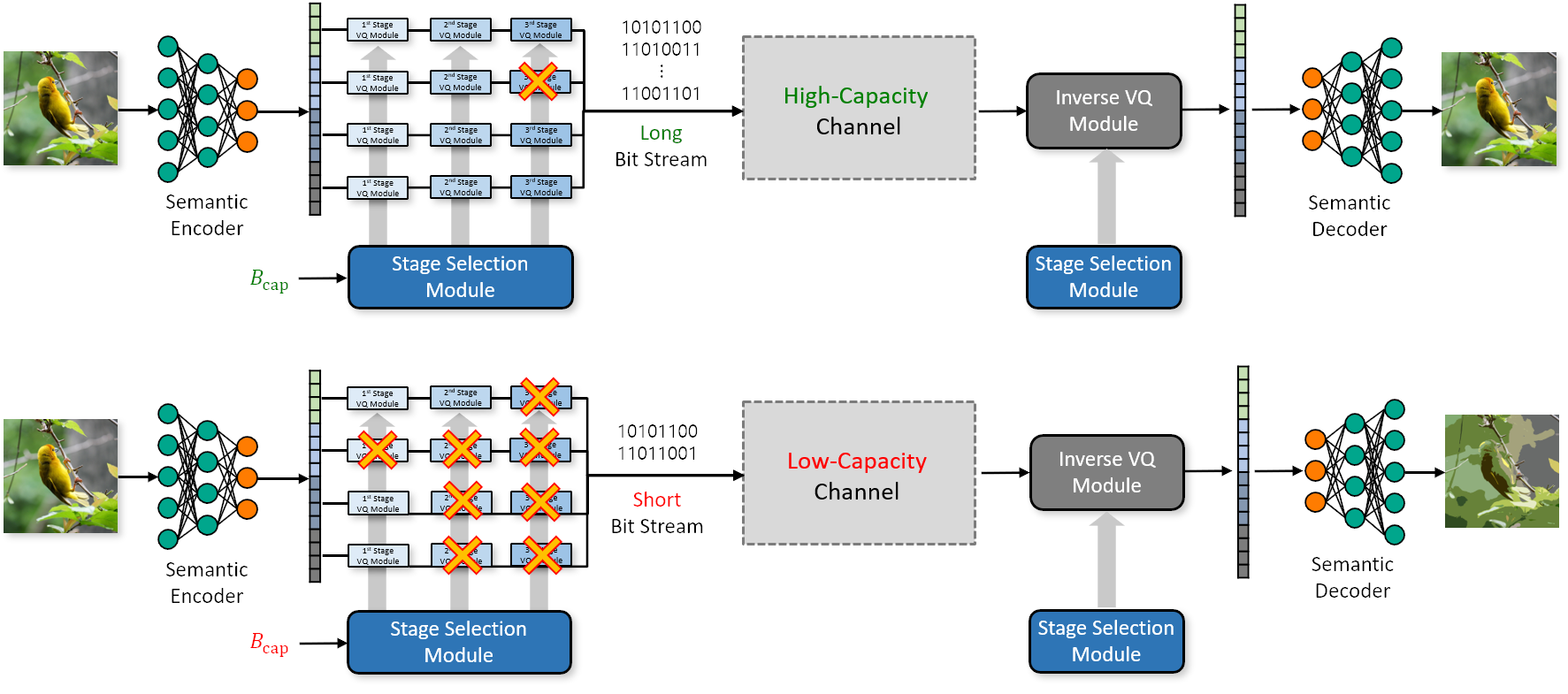,width=15cm}}
    \caption{An illustration of rata-adaptive operation via stage selection in the proposed MSVQ-SC framework.}\vspace{-5mm}
    \label{fig:ModuleSelect}
\end{figure*}

The aim of our strategy is to enable rate-adaptive SC, dynamically adjusting the number of transmitted bits to meet the rate constraint specified by $B_{\rm cap}$. 
Recall that the number of bits for representing the output of the VQ module for the $i$-th sub-vector at stage $t$ is given by $B_i^{(t)}$.
Therefore, if the number of stages used to quantize the $i$-th sub-vector is set as $T_i \leq T_{\rm max}$, the total number of bits representing the semantic feature vector ${\bf z}$ is given by
\begin{align}\label{eq:B_tot}
    B_{\rm tot} = \sum_{i=1}^N \sum_{t=1}^{T_i} B_i^{(t)},
\end{align}
This expression implies that $B_{\rm tot}$ can be dynamically adjusted by varying the number of stages applied to each sub-vector. For example, if $T_i$ is set to its maximum value (i.e., $T_i=T_{\rm max}$) for every sub-vector, the resulting bit overhead is given by $\sum_{i=1}^N \sum_{t=1}^{T_{\rm max}} B_i^{(t)}$ bits with the best performance. Conversely, if $T_i$ is set to its minimum value (i.e., $T_i = 0$), no quantization is performed, yielding a total of 0 bits with the worst performance. Thus, by appropriately selecting the number of stages across sub-vectors, the bit overhead can range from $0$ to $\sum_{i=1}^N \sum_{t=1}^{T_{\rm max}} B_i^{(t)}$ with a granularity of $B_i^{(t)}$ bits. This approach offers fine-grained rate control that is easily implemented via a simple switching operation. Motivated by these considerations, we propose a stage selection strategy that determines the number of stages for each sub-vector under the constraint $B_{\rm tot} \leq B_{\rm cap}$. 
Our stage selection strategy is illustrated in Fig.~\ref{fig:ModuleSelect}.

\subsection{Optimization Problem for Stage Selection}

In the stage selection approach, an arbitrary selection of stages for $N$ sub-vectors may lead to a substantial degradation in task performance, particularly at lower rates where fewer stages are used. Because each module contributes differently to the fidelity of the reconstructed data and overall task performance, a naive or random selection might omit the modules critical for preserving semantic content. 

To ensure robust task performance across varying rates, it is essential to prioritize the selection of stages based on their relative contribution to overall task quality. Motivated by this observation, we formulate an optimization problem to determine the optimal stage selection under the constraint on the total number of bits.
Our problem is formulated as 
\begin{align}\label{eq: Loss Minimization prob}
    \min_{\{T_i\}_i} ~&\sum_{i} \mathcal{E}_i\big({\bf x},\hat{\bf z}_{i}^{(T_i)}\big)  \\
    \text{s.t.}~~&\sum_{i=1}^N \sum_{t=1}^{T_i} B_i^{(t)} \leq B_{\rm cap},~~0\leq T_i \leq T_{\rm max},\nonumber
\end{align}
where $\mathcal{E}_i\big({\bf x},\hat{\bf z}_{i}^{(T_i)}\big)$ is a performance metric measured for the input data ${\bf x}$ and the $i$-th quantized sub-vector $\hat{\bf z}_{i}^{(T_i)}$ when using the VQ modules up to stage $T_i$.

We consider a metric that evaluates the stage selection quality based on a task loss. This metric evaluates the impact of each quantization module on the overall task loss when included or excluded. 
It enables the system to rank modules by their relevance, ensuring that those with greater contributions to task loss reduction are prioritized under the bit constraint. To compute this metric, after the joint training process in Sec.~\ref{Joint_Training}, we assess a marginal task loss by assuming that the $i$-th sub-vector is quantized using only up to stage $T_i$, while other sub-vectors are quantized using all $T_{\rm max}$. This assumption isolates the marginal contribution of the stage selection for the $i$-th sub-vector without being influenced by the selection in other sub-vectors. Following this strategy,  for an image reconstruction task, the marginal loss is defined as 
\begin{align}\label{eq: Task loss}
    \mathcal{E}\big({\bf x},\hat{\bf z}_{\setminus i}^{(T_i)}\big)  =\|{\bf x} - f_{\bm \phi}(\hat{\bf z}_{\setminus i}^{(T_i)})\|^2,
\end{align}
where $\hat{\bf z}_{\setminus i}^{(T_i)}$ denotes the output of the overall quantization module in which all sub-vectors except the $i$-th sub-vector use all $T_{\rm max}$ stages, while the $i$-th sub-vector uses only up to stage $T_i$. Notably, this metric can be adapted depending on the target task. For example, in a classification scenario, it may be computed using the cross-entropy loss. This flexibility ensures that the proposed framework is applicable across various types of SC tasks.

Although the marginal loss in \eqref{eq: Task loss} is a direct measure for minimizing the task loss, computing this metric requires the knowledge of the decoder $f_{\bm \phi}(\cdot)$, which is often not available at the transmitter. To circumvent such impractical requirement, we approximate the marginal loss using its empirical estimate computed from the training data, defined as
\begin{align}\label{eq: Average loss}
   \mathbb{E}_{{\bf x}}\big[ \mathcal{E}_i\big({\bf x},\hat{\bf z}_{i}^{(T_i)}\big) \big] = \frac{1}{|\mathcal{D}|}\sum_{{\bf x}^\prime \in\mathcal{D}} \mathcal{E}\big({\bf x}^\prime,\hat{\bf z}_{\setminus i}^{(T_i)}\big).   
\end{align}
We refer to this metric as an {\em empirical} marginal loss.
With this approximation, our optimization problem becomes
\begin{align}\label{eq:Problem2}
    \min_{\{T_i\}_i}~&\sum_{i} \mathbb{E}_{{\bf x}}\big[ \mathcal{E}_i\big({\bf x},\hat{\bf z}_{i}^{(T_i)}\big) \big] \\
    \text{s.t.}~~&\sum_{i=1}^N \sum_{t=1}^{T_i} B_i^{(t)}   \leq B_{\rm cap},~~0\leq T_i \leq T_{\rm max}. \nonumber
\end{align}
Because the objective function of the reformulated problem can be computed during offline training, the resulting values can be shared between the transmitter and the receiver via a lookup table comprising $NT_{\rm max}$ values. Based on this lookup table, both the transmitter and the receiver can independently solve the same stage selection problem, ensuring that they arrive at identical selection results without incurring extra communication overhead.

\subsection{Proposed Stage Selection Algorithm}\label{Sec:Selection_Algorithm}
The stage selection problem in \eqref{eq:Problem2} is a discrete optimization problem. Suppose that the number of quantization bits allocated to each sub-vector is identical\footnote{This assumption may not hold depending on the quantization bit allocation strategy. In this work, we intentionally allocate similar quantization bits to all stages for each sub-vector so that the condition $B_i^{(t)}=B_i$ for all $t\in[T_{\rm max}]$ holds approximately.} across all stages, i.e., $B_i^{(t)}=B_i$ for all $t\in[T_{\rm max}]$. In this case, an incremental allocation algorithm is known to yield an optimal solution \cite{Fox}, provided that the objective function is strictly convex and monotonically decreasing with respect to the number of selected stages. 

Motivated by this result, we adopt an incremental allocation algorithm to solve the stage selection problem in \eqref{eq:Problem2}. In this algorithm, we iteratively increase the stage assignment variable $T_i$ by one for the sub-vector that yields the largest decrease in the objective. Specifically, at each iteration, we select the best index $i^*$ of the sub-vector, determined by
\begin{align}\label{eq:i_select}
    i^* = \underset{i: T_i< T_{\rm max}}{\rm argmax}~   \frac{\mathbb{E}_{{\bf x}}\big[ \mathcal{E}_i\big({\bf x},\hat{\bf z}_{i}^{(T_i)}\big) \big] - \mathbb{E}_{{\bf x}}\big[ \mathcal{E}_i\big({\bf x},\hat{\bf z}_{i}^{(T_i+1)}\big) \big]}{B_i^{(T_i+1)}}.
\end{align}
Note that the search is restricted to the sub-vectors whose stage assignments have not yet reached $T_{\rm max}$. 
This iterative process continues until the total bit overhead reaches its limit $B_{\rm cap}$. Through this procedure, the algorithm effectively captures the relative priority of each module based on its impact on the average performance loss. A VQ module is considered more important if assigning it an additional stage leads to a larger reduction in the objective function. As a result, this importance-aware selection enables the system to prioritize and transmit the most critical features, achieving effective performance under a given bit overhead constraint.
Our stage selection algorithm is summarized in Algorithm~\ref{alg: MStage Selection}.

A promising feature of the proposed algorithm is that it can be executed offline after the training phase. Specifically, the empirical marginal loss in \eqref{eq: Average loss} can be computed using the learned semantic encoder, decoder, and VQ modules. Based on this computation, the stage selection order can be pre-determined and stored at both the transmitter and receiver. As a result, no additional signaling or computational overhead is required during online communication.

{\small \RestyleAlgo{ruled}
    \SetKwComment{Comment}{/* }{ */}
    \begin{algorithm}[t]\label{alg: MStage Selection}
    \caption{Proposed Stage Selection Algorithm}
    \setstretch{1.2}
    \textbf{1. Initialization:}\\
    $T_i=0,~\forall i\in [N]$;\\

    \textbf{2. Stage selection process:}\\
    \While{$\sum_{i=1}^N \sum_{t=1}^{T_i} B_i^{(t)} < B_{\rm cap}$}{
    
    Choose $i^*$ from \eqref{eq:i_select}; \\
    $T_{i^{*}}\leftarrow T_{i^{*}}+1$; \\

    }
    \end{algorithm}}

\begin{figure}[t]
    \centering
    {\epsfig{file=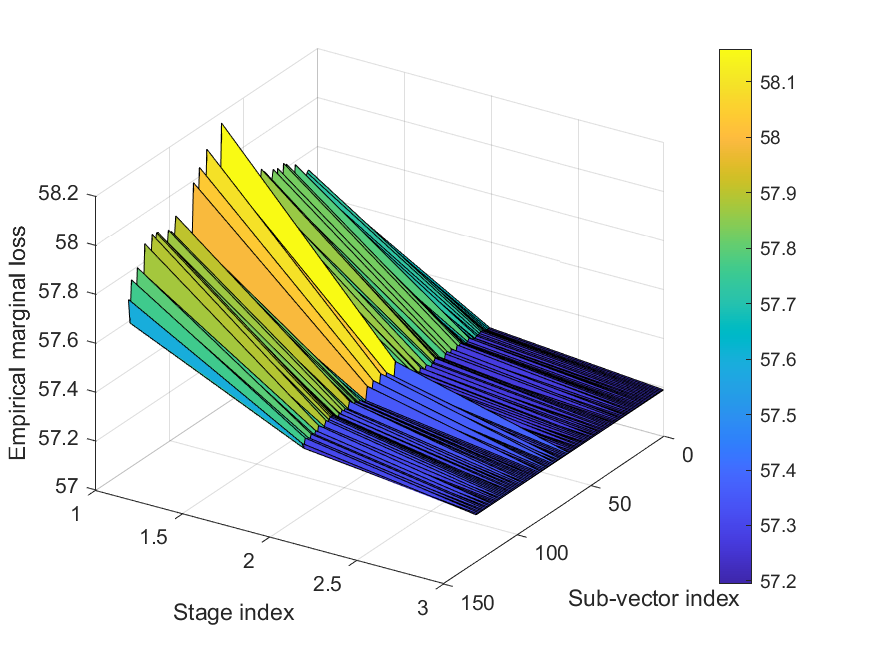,width=9.5cm}}
    \caption{Numerical results for the average marginal task loss based on the quantization error.}\vspace{-3mm}
    \label{fig: empirical expectation}
\end{figure}

{\bf Remark 2 (Optimality of Stage-Selection Algorithm):}
Our stage selection algorithm in Algorithm~\ref{alg: MStage Selection} is guaranteed to be optimal only when the empirical marginal loss $\mathbb{E}_{{\bf x}}\big[ \mathcal{E}_i\big({\bf x},{\bf z}_{q,i}^{(T_i)}\big) \big]$ in \eqref{eq:Problem2} is strictly convex and monotonically decreasing \cite{Fox}. Strictly speaking, this property may not always hold, as it can depend on various factors such as the input data distribution, the target task, and the design of the encoder-decoder network. Nevertheless, in most practical scenarios, the task loss is expected to be monotonically decreasing with $T_i$ since the difference between the semantic feature vector ${\bf z}$ and its quantized representation typically reduces as the number of VQ stages increases. In addition, the task loss often exhibits an approximately convex trend, because the performance improvement becomes progressively less sensitive to further reductions in quantization error.
To validate these conjectures, Fig.~\ref{fig: empirical expectation} plots the empirical marginal loss for the scenario described in Sec.~VI. The numerical results demonstrate that our loss exhibits a convex-like trend and a monotonically decreasing property.
This numerical evidence not only supports our conjecture regarding the approximate convexity and monotonicity of the empirical marginal loss, but also demonstrates the effectiveness of our stage selection algorithm in Algorithm~\ref{alg: MStage Selection} for solving the problem in \eqref{eq:Problem2}, despite the lack of a formal guarantee of optimality.

\section{Enhancement of The Proposed MSVQ-SC Framework via Entropy Coding}\label{Sec:Per_VQ_module_2}


Recall that in the proposed MSVQ-SC framework, the transmitter conveys the indices of the codeword vectors determined by the selected VQ modules. 
Typically, when using learned VQ codebooks as in the proposed framework, the selection frequency of codeword vectors in $\mathcal{C}_i^{(t)}$ tends to follow nonuniform distributions \cite{ECVQ}. This observation suggests that the bit overhead can be further reduced by applying entropy coding when representing the selected codeword vector at each stage. Inspired by this observation, in this section, we further enhance the proposed MSVQ-SC framework by incorporating the entropy coding into each VQ module. We refer to this modified framework as an entropy-constrained MSVQ-SC (EC-MSVQ-SC) framework.

\subsection{Incorporating Entropy Coding into MSVQ-SC}
Let ${\bf c}_{i,k}^{(t)} \in \mathbb{R}^D$ be the $k$-th codeword vector in the codebook $\mathcal{C}_i^{(t)}$. Then, the codebook for the $i$-th sub-vector in stage $t$ is expressed as 
\begin{align}
    \mathcal{C}_i^{(t)} = \big\{ {\bf c}_{i,1}^{(t)}, {\bf c}_{i,2}^{(t)}, \ldots, {\bf c}_{i,K_i^{(t)}}^{(t)}\big\}.
\end{align}
To apply entropy coding, after the joint training process, we measure the probability mass function of ${\bf c}_{i,k}^{(t)}$ by using the training data as follows:
\begin{align}\label{eq:PMF_EC}
   p_{i,k}^{(t)}  &= \mathbb{P}\big[\tilde{\bf z}_i^{(t)} = {\bf c}_{i,k}^{(t)}  \big] \nonumber \\
   &= \frac{1}{|\mathcal{D}|}\sum_{{\bf x}^\prime \in\mathcal{D}} \mathbb{I}\Big[ Q_{\bar{\mathcal{C}}_i^{(t)}}^{(t)}\big(f_{{\bm \theta}}({{\bf x}})\big) = {\bf c}_{i,k}^{(t)}\Big],
\end{align}
where $Q_{\bar{\mathcal{C}}^{(t)}_i}^{(t)}({\bf z})=\tilde{\bf z}_i^{(t)}$ denotes the output of the VQ modules up to stage $t$ for the $i$-th sub-vector, and $\bar{\mathcal{C}}_i^{(t)}$ is a set of the VQ codebooks up to stage $t$ for the $i$-th sub-vector, i.e., 
\begin{align}
    \bar{\mathcal{C}}_i^{(t)} = \{\mathcal{C}_i^{(t^\prime)}\}_{t^\prime \leq t},~\forall t\in [T_{\rm max}].
\end{align}
The entropy coding schemes, such as Huffman coding \cite{EC1} or arithmetic coding \cite{EC2}, can be constructed based on the probability mass function for $\{p_{i,k}^{(t)}\}_{k=1}^{2^{B_{i}^{(t)}}}$. As a result, the average number of the bits required to convey the information of $\tilde{\bf z}_i^{(t)}$ after the entropy coding is given by 
\begin{align}\label{eq:B_EC_i}
    \mathbb{E}\big[{B}_{{\rm EC},i}^{(t)}\big] = - \sum_{k=1}^{K_i^{(t)}} p_{i,k}^{(t)}  \log_2 p_{i,k}^{(t)}.
\end{align}
This average overhead satisfies $ \mathbb{E}\big[{B}_{{\rm EC},i}^{(t)}\big]  \leq B_i^{(t)}$, with equality holding only when the selection probability is uniform, i.e., $p_{i,k}^{(t)}={1}/K_i^{(t)}$ for all $k\in [K_i^{(t)}]$.

By taking into account the use of the entropy coding, we also modify the original quantization rule in \eqref{eq:VQ_quantize} to reflect a trade-off between distortion and rate, as done in \cite{OriginalECVQ}. This modification allows the model to favor codeword vectors that are not only close to the input residual vector but also statistically more probable. The modified quantization rule is given by 
\begin{align}\label{eq:Modified quantize}
    \tilde{\mathbf{z}}_{i}^{(t)}=\argmin_{\mathbf{c}_k \in \mathcal{C}_i^{(t)}}~\lambda^{(t)} \Vert\mathbf{r}_i^{(t-1)} - \mathbf{c}_k \Vert^2-\log_2 p_{i,k}^{(t)},
\end{align}
where $\lambda^{(t)}$ serves as a hyperparameter that governs the trade-off between distortion minimization and entropy regularization. This formulation encourages the selection of codewords that are not only close to the residual vector but also likely to occur, leading to more efficient entropy coding.

By considering the rate-distortion optimality, the semantic encoder, decoder, and VQ modules need to be jointly trained with the codeword distribution $p_{i,k}^{(t)}$. To make this distribution learnable, we model $p_{i,k}^{(t)}$ in \eqref{eq:Modified quantize} using trainable parameters $\{w_{i,k}^{(t)}\}_{\forall i, k }$ as follows \cite{ECVQ}:
\begin{align}\label{eq:codeword distribution}
    \hat{p}_{i,k}^{(t)} = \frac{e^{-w_{i,k}^{(t)}}}{\sum_{j=1}^{K_i^{(t)}} e^{-w_{i,j}^{(t)}}},
\end{align}
where each $w_{i,k}^{(t)}$ is a trainable parameter associated with the $k$-th codeword of the $i$-th sub-vector in stage $t$. This parameterization allows the quantizer to adapt its codeword distribution to the underlying data distribution for an improved rate efficiency. Based on this strategy, we jointly train the semantic encoder, decoder, VQ modules, and codeword distribution using the joint loss function suggested in \cite{ECVQ}. Details of the joint training process can be found in \cite{ECVQ}.

\subsection{Modified Rate-Adaptive Communication Strategy}\label{Optimization problem w EC}
Unlike in Sec.~\ref{Stage Selection}, where a fixed-length representation is assumed for codeword indices, the integration of entropy coding into the proposed MSVQ-SC framework results in variable-length bit representations. Specifically, since entropy coding assigns shorter bit sequences to more frequent codewords and longer sequences to less frequent ones, the number of bits required to represent each selected codeword varies depending on its probability within each VQ module.
As a consequence, the total bit length is no longer fixed but varies depending on the specific codewords selected from multiple VQ modules. 

To accommodate this variability, we modify the rate-adaptive communication strategy in Sec.~\ref{Stage Selection} by accounting for the entropy-induced bit-lengths associated with each quantization stage. To this end, the stage selection problem in \eqref{eq:Problem2} must be reformulated by considering the bit budget constraint that reflects the variable-length property of the entropy coding. Recall that the average bit length of the codeword representation corresponding to the $i$-th sub-vector in stage $t$ is given as in \eqref{eq:B_EC_i}. The modified problem to optimize the stage selection is given by 
\begin{align}\label{eq:Problem_EC}
    \min_{\{T_i\}_i}~&\sum_{i} \mathbb{E}_{{\bf x}}\big[ \mathcal{E}_i\big({\bf x},\hat{\bf z}_{i}^{(T_i)}\big) \big] \\
    \text{s.t.}~~&\sum_{i=1}^N \sum_{t=1}^{T_i} \bar{B}_{{\rm EC},i}^{(t)} \leq B_{\rm cap},~~0\leq T_i \leq T_{\rm max}, \nonumber
\end{align}
where $ \bar{B}_{{\rm EC},i}^{(t)}$ denotes the average number of bits assigned to the $i$-th sub-vector position at stage $t$, which is empirically computed using the training dataset based on the selected entropy coding method. The stage selection problem in \eqref{eq:Problem_EC} is also a discrete optimization problem. The only difference with the problem in \eqref{eq:Problem2} lies in the fact that the bit constraint is applied with respect to the average  bit length $\bar{B}_{{\rm EC},i}^{(t)}$, instead of the quantization bit $B_i^{(t)}$.

Similar to the stage selection algorithm in Sec.~\ref{Sec:Selection_Algorithm}, we solve the above problem by employing an incremental allocation algorithm introduced in \cite{Fox}. Specifically, we select the best index $i^*$ of the sub-vector, determined by
\begin{align}\label{eq:i_select w bit length}
    i^* = \underset{i: T_i< T_{\rm max}}{\rm argmax}~    \frac{ \mathbb{E}_{{\bf x}}\big[\mathcal{E}_i\big({\bf x},\hat{\bf z}_{i}^{(T_i)}\big)\big]  -  \mathbb{E}_{{\bf x}}\big[\mathcal{E}_i\big({\bf x},\hat{\bf z}_{i}^{(T_i+1)}\big)\big]}{{ \bar{B}_{{\rm EC},i}^{(T_i+1)} }},
\end{align}
until the cumulative number of average bits consumed by all selected modules does not exceed the total communication budget. Note that this algorithm can be performed offline after computing the probability mass function from \eqref{eq:PMF_EC}. As a result, the stage selection order can be saved at both the transmitter and receiver, implying that no extra signaling overhead is needed for sharing the stage selection.

The above stage selection algorithm is not guaranteed to be optimal because the average bit lengths after entropy coding may differ across stages depending on the codeword distributions and selected codewords, even if the objective function is strictly convex and monotonically decreasing with respect to the number of selected stages.
Nevertheless, this algorithm offers a practical alternative to solve the stage selection problem without incurring a significant computational burden.
Additionally, the algorithm allows the system to prioritize fine-grained encoding of the most essential components by considering the expected loss reduction with respect to the cost of the increase in the average bit length.

\vspace{1mm}
{\bf Remark 3 (Shared Codebook for Memory Reduction)}: 
The number of codebook parameters required for the proposed MSVQ-SC and EC-MSVQ-SC frameworks is given by 
\begin{align}
    D\sum_{i=1}^N\sum_{t=1}^{T_{\rm max}}K_i^{(t)},
\end{align}
which can become prohibitively large when $D$, $K_i^{(t)}$, and $N$ are large.
To mitigate excessive memory requirements, we introduce a simple modification in which VQ codebooks are shared across multiple VQ modules that use the same codebook size. Specifically, the $N$ sub-vectors are divided into $G$ groups such that all VQ modules within the same group are constrained to have the same codebook. With this modification, the number of the codebook parameters becomes 
\begin{align}
    DG\sum_{t=1}^{T_{\rm max}}K_i^{(t)}.
\end{align}
This result implies that the number of parameters can be significantly reduced by setting $G$ to a small value. Although each shared codebook must learn more diverse distributions as $G$ decreases, the resulting performance degradation is expected to be not significant, as will be shown in Sec. VI.

\section{Simulation Results}\label{Sec:Simulation Results}

In this section, we evaluate the superiority of the proposed MSVQ-SC and EC-MSVQ-SC frameworks compared to existing digital SC methods. 

\begin{figure*}[t]
    \centering
    \subfloat[PSNR]{\includegraphics[width=0.333\textwidth]{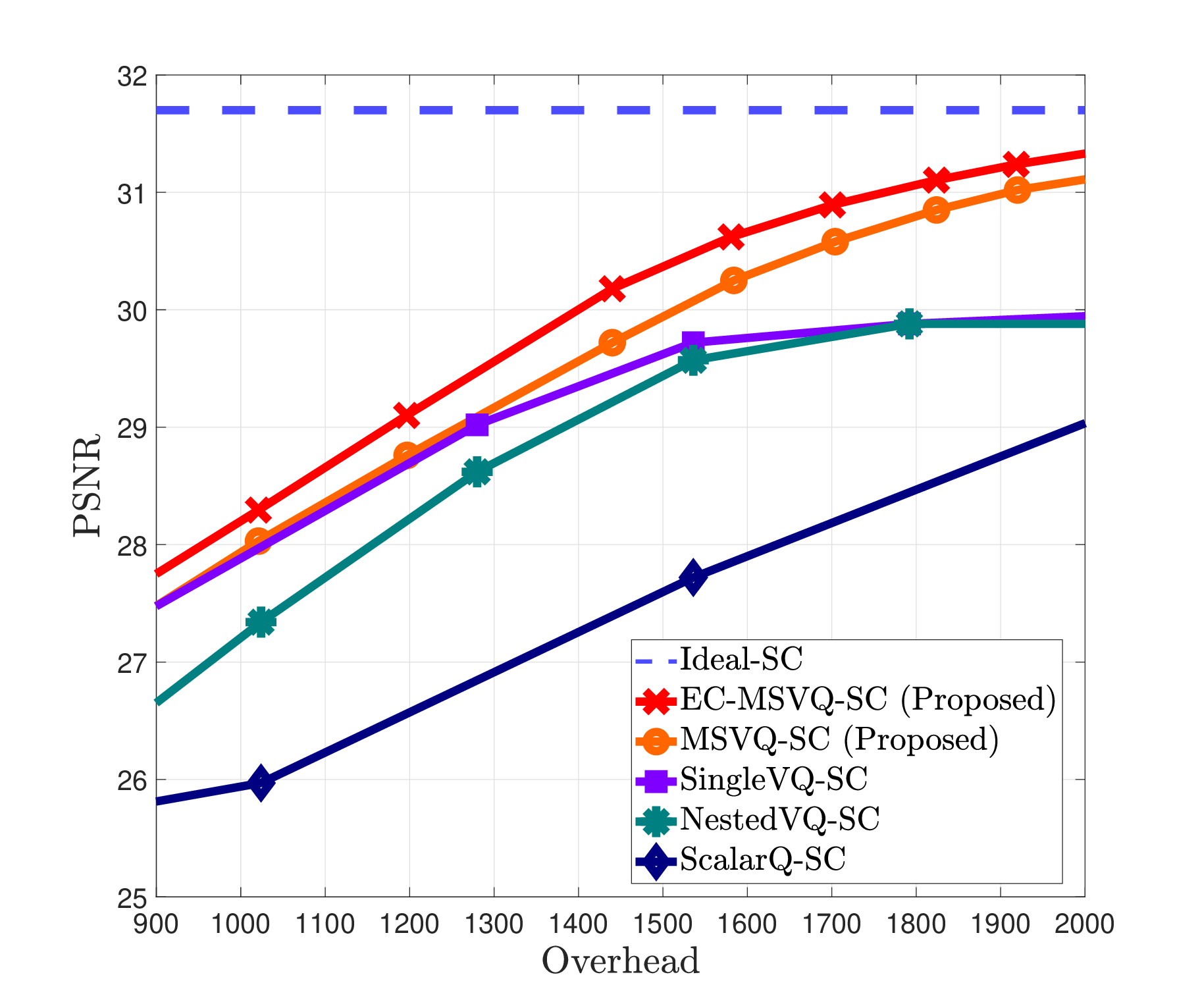}\label{fig:psnr_6bit}}
    \hfill
    \subfloat[LPIPS]{\includegraphics[width=0.333\textwidth]{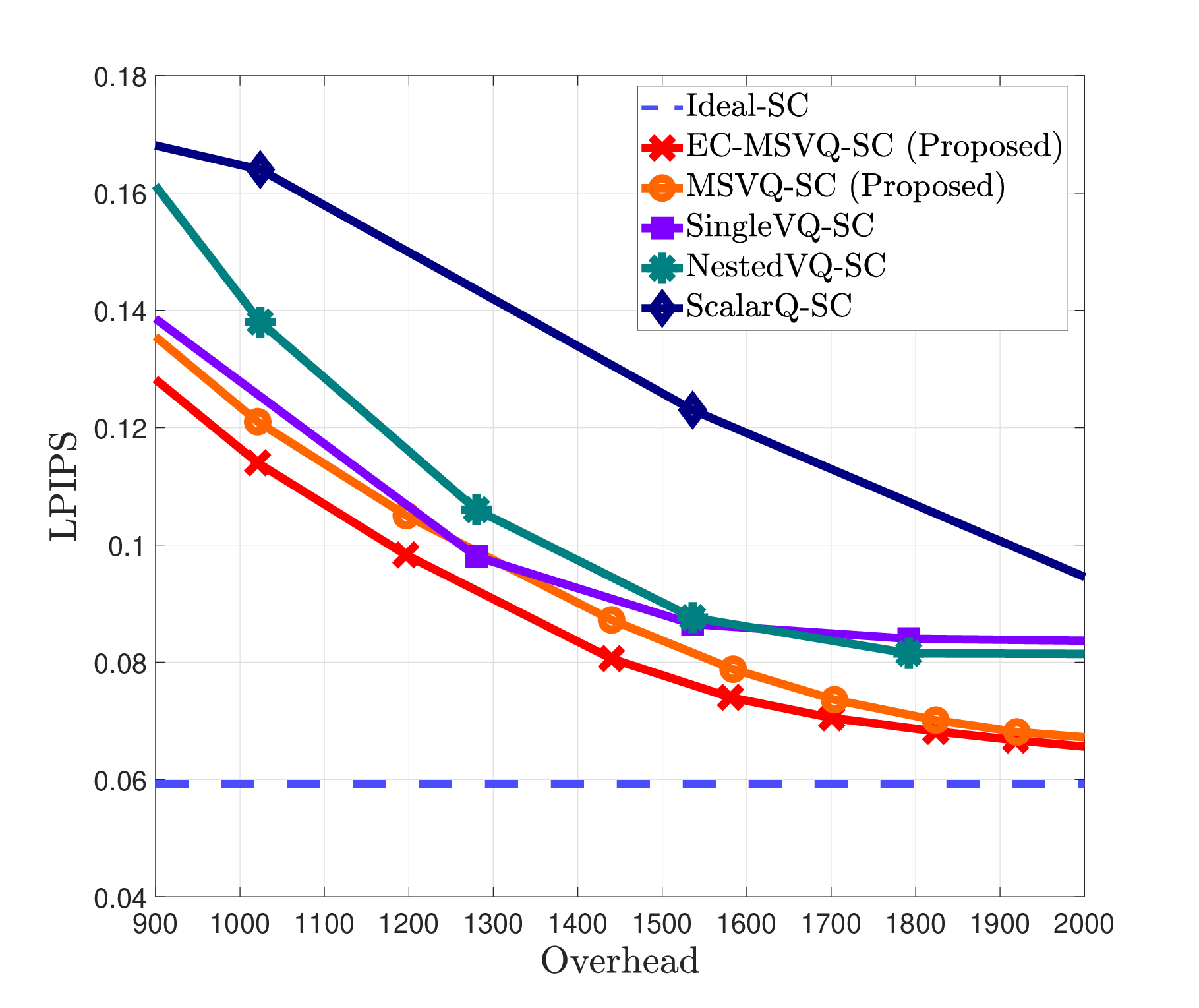}\label{fig:LPIPS_6bit}}
    \hfill
    \subfloat[SSIM]{\includegraphics[width=0.333\textwidth]{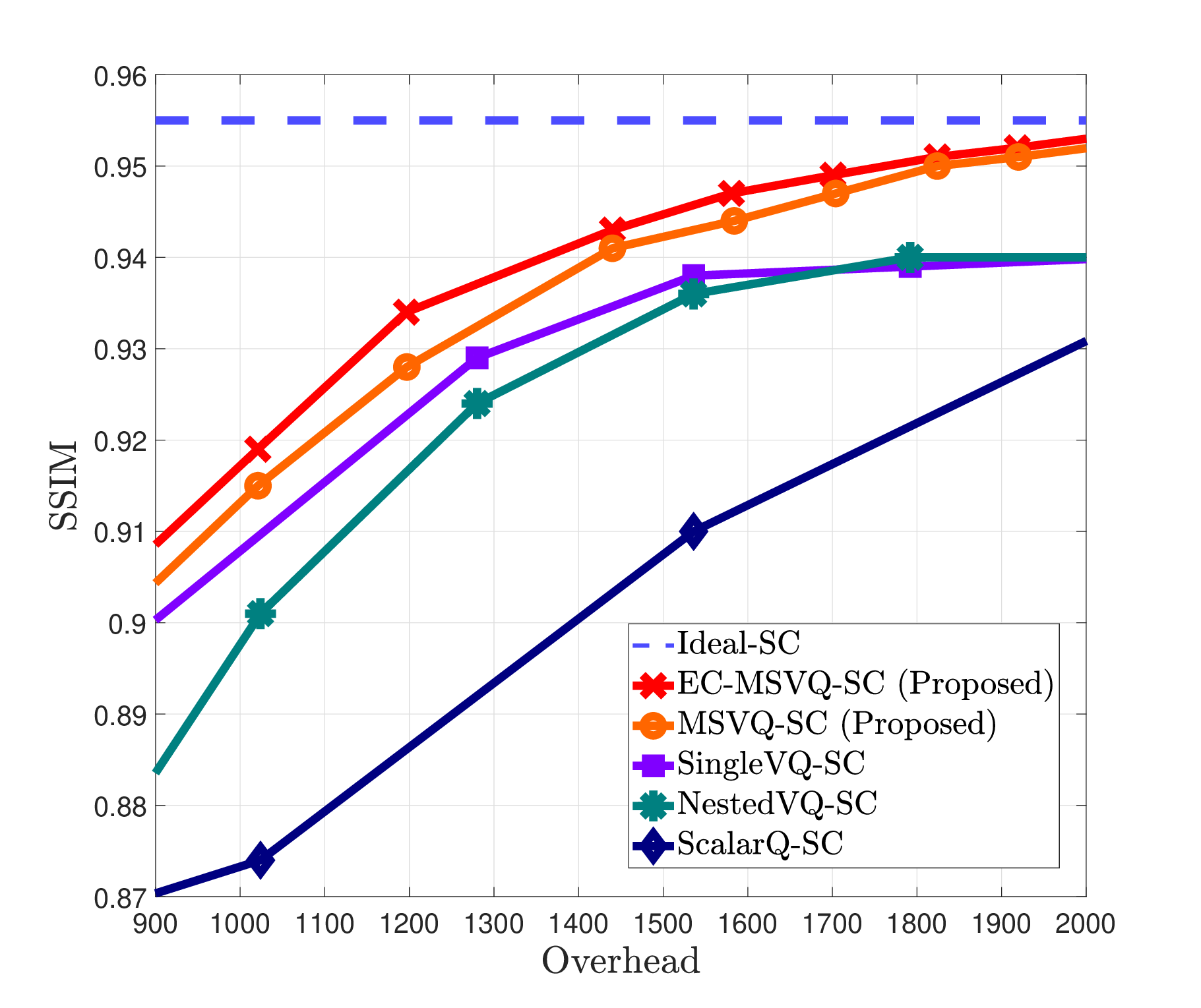}\label{fig:SSIM_6bit}}
    \caption{Comparison of the performances of various digital SC frameworks under the finite-capacity scenario.}
    \label{fig:ReconstvsOverhead}
\end{figure*}

\subsection{Experiment Setup}\label{Experiment Setup}

We consider an image reconstruction task using the CIFAR-10 dataset. The CIFAR-10 dataset contains 60,000 color images of size $32 \times 32 \times 3$ across 10 classes, with 50,000 images used for training and 10,000 for testing. The encoder for CIFAR-10 consists of five convolutional layers: two downsampling layers with 64 and 128 output channels (kernel size $5\times5$, stride 2), followed by two convolutional layers with 128 channels (kernel size $5\times5$, stride 1), and a final layer producing 8 channels (kernel size $3\times3$, stride 1). Each layer, except the last, is followed by a PReLU activation. The decoder is constructed symmetrically with five transposed convolutional layers: one layer mapping 8 to 128 channels (kernel size $3\times3$, stride 1), two layers with 128 channels (kernel size $5\times5$, stride 1), one upsampling layer to 64 channels (kernel size $5\times5$, stride 2, output padding 1), and a final layer that reconstructs the RGB image with 3 output channels (kernel size $5\times5$, stride 2, output padding 1). PReLU activations are used after all but the last layer.


For performance comparison, we consider the following SC frameworks: 
\begin{itemize}
    \item {\bf Ideal-SC}: This is an ideal SC framework without any quantization error (i.e., $\hat{\bf z}={\bf z}$). The semantic encoder and decoder are trained without quantization effects. This method offers a performance upper bound for all the considered digital SC frameworks.
    
    \item {\bf EC-MSVQ-SC (Proposed)}: This is the proposed MSVQ-SC framework with entropy coding with a modified stage selection strategy described in Sec. V.  Unless specified otherwise, we set $\lambda^{(1)}=2000$, $\lambda^{(2)}=5000$, and $\lambda^{(3)}=10000$.
    
    \item {\bf MSVQ-SC (Proposed)}: This is the proposed MSVQ-SC framework without entropy coding along with the stage selection method in Sec. IV. 
    
    \item {\bf ScalarQ-SC}: This is an existing digital SC framework using the $\mu$-law scalar quantization. This is a post quantization method which does not  consider the joint training of the quantization module with the semantic encoder and decoder. 
    
    \item {\bf SingleVQ-SC \cite{VQ0}}: This is an existing digital SC framework using the single-stage VQ framework in \cite{VQ0}. This method jointly trains the semantic encoder and decoder along with VQ codebooks. Since this framework does not support a rate-adaptive operation, it requires multiple pairs of the semantic encoder-decoder to support multiple transmission rates.   
    
    \item {\bf NestedVQ-SC \cite{nested}}: This is an existing digital SC framework using the single-stage VQ framework in \cite{nested}. This method jointly trains the semantic encoder and decoder along with VQ codebooks. Unlike SingleVQ-SC, this framework supports multiple transmission rates by introducing a nested codebook design. Therefore, this framework requires only a single pair of the semantic encoder and decoder. 
    
\end{itemize}

The key parameters of the proposed framework are set as follows: the number of quantization stages $T_{\rm max}=3$, the latent vector dimension $M=512$, the sub-vector dimension $D=4$, the number of sub-vectors $N=128$, and the commitment loss coefficient $\beta=0.25$. For the first stage, we assign an 8-bit quantization codebook to the 64 sub-vectors with the highest variance and a 6-bit codebook to the remaining 64 sub-vectors. For the second stage, 7-bit and 5-bit codebooks are assigned to the high-variance and low-variance sub-vectors, respectively. For the third stage, 6-bit and 4-bit codebooks are assigned in the same manner. Unless otherwise specified, each of the 128 sub-vectors in each stage is assigned a different quantization codebook. The loss weights for stages 1, 2, and 3 are set to $w_1=0.2$, $w_2=0.2$, and $w_3=1$, respectively. We train the encoder, quantization modules, and decoder using the sequential training strategy described in Sec.~\ref{Joint_Training}, employing the Adam optimizer with a learning rate of $10^{-4}$ for a total of 30 epochs.

\subsection{Evaluation under Finite-Capacity Scenarios}\label{Image reconstruction}

To evaluate image reconstruction performance in various aspects, we consider three performance metrics. (i) Peak signal-to-noise ratio (PSNR), which is widely used in the literature to measure the per-pixel difference between original and reconstructed images, (ii) Structural similarity index (SSIM), which considers luminance, contrast, and structural information to better reflect structural preservation, and (iii) Learned perceptual image patch similarity (LPIPS), which leverages features from a pre-trained VGG network to measure perceptual similarity.

In Fig.~\ref{fig:ReconstvsOverhead}, we compare the performance of various digital SC frameworks under the finite-capacity scenario using three metrics: PSNR in Fig.~\ref{fig:ReconstvsOverhead}(a), LPIPS in Fig.~\ref{fig:ReconstvsOverhead}(b), and SSIM in Fig.~\ref{fig:ReconstvsOverhead}(c). In particular, Fig.~\ref{fig:ReconstvsOverhead}\subref{fig:psnr_6bit} shows that the proposed frameworks achieve superior reconstruction fidelity compared to all baselines, with especially notable gains in the medium-to-high bit budget regimes. This result confirms that the multi-stage design more effectively captures task-relevant semantic features and reduces reconstruction errors. Fig.~\ref{fig:ReconstvsOverhead}\subref{fig:LPIPS_6bit} further demonstrates that the proposed frameworks consistently achieve lower LPIPS scores across nearly all bit constraints, reflecting both improved perceptual quality and better preservation of semantic information. Finally, Fig.~\ref{fig:ReconstvsOverhead}\subref{fig:SSIM_6bit} indicates that the proposed frameworks achieve higher SSIM values than the baselines, highlighting their ability to preserve structural information and maintain spatial integrity in reconstructed data. This property is particularly advantageous in applications such as object recognition and downstream inference, where structural consistency is critical.

It is worth noting that although SingleVQ-SC provides the most competitive baseline performance, it requires multiple encoder-decoder pairs, whereas the proposed frameworks employ only a single pair. This demonstrates that the proposed frameworks achieve rate adaptivity without sacrificing performance by judiciously selecting the most effective VQ modules for a given bit budget. Moreover, the performance gains of EC-MSVQ-SC over MSVQ-SC confirm that the incorporation of entropy coding brings additional transmission efficiency by exploiting the non-uniform codeword distributions naturally present in the learned VQ codebooks.
Taken together, the consistent improvements across PSNR, LPIPS, and SSIM confirm that the superiority of the proposed frameworks is not limited to a single metric. Instead, the results highlight its robustness and adaptability across diverse bit budgets, while the compact multi-stage codebook ensures reduced computational complexity. These advantages establish the proposed frameworks as a practical and scalable solution for rate-adaptive SC.
\begin{figure}[t]
    \centering
    {\epsfig{file=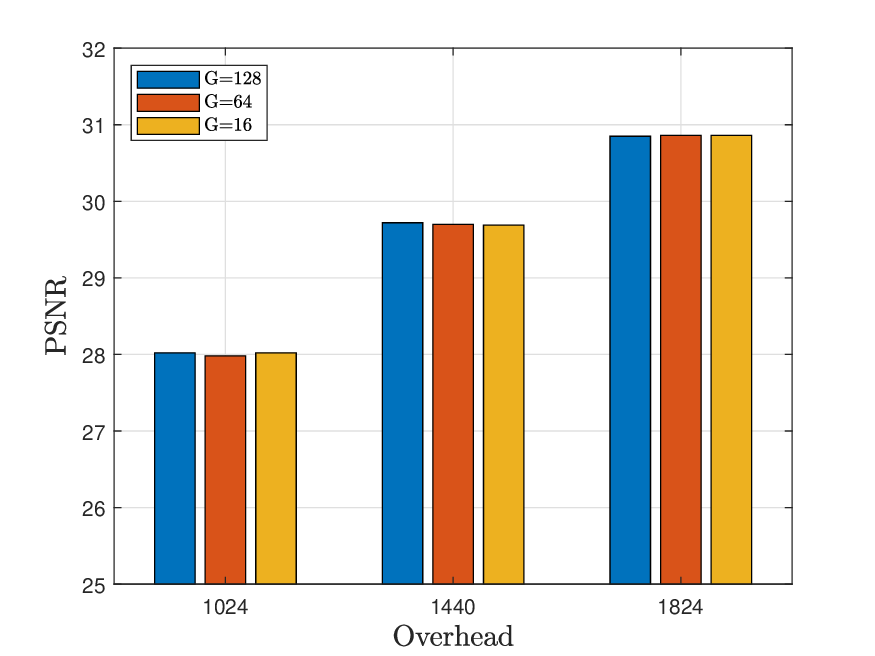,width=9cm}}
    \caption{Comparison of the PSNR performance of the proposed MSVQ-SC framework with different numbers of sub-vector groups.}\vspace{-3mm}
    \label{fig:GroupingvsOverhead}
\end{figure}

We now explore the effectiveness of the shared codebook strategy introduced in {\bf Remark 3}, which is designed to reduce the memory requirements of the proposed frameworks. To this end, in Fig.~\ref{fig:GroupingvsOverhead}, we compare the PSNR performance of the proposed MSVQ-SC framework under the finite-capacity scenario for different numbers of groups, $G$.
The results show that although the performance of the framework slightly decreases as $G$ becomes smaller, the degradation remains marginal. Meanwhile, the memory requirement of the framework with 
$G=16$ is reduced by a factor of eight compared to the case with $G=128$. These findings confirm that the shared codebook strategy provides a highly effective means of mitigating memory burden while preserving reconstruction performance.



\begin{table}[t]
\centering
\caption{PSNR performance of the proposed MSVQ-SC framework with different bit allocation strategies.}
\label{tab:psnr_bitbudget}
\begin{tabular}{cccccc}
\hline
\textbf{Bit overhead} & \textbf{896} & \textbf{1024} & \textbf{1200} & \textbf{1920} & \textbf{2088} \\
\hline
{\bf Type I} & $\mathbf{27.46}$ & $\mathbf{28.02}$ & $\mathbf{28.70}$ & $\mathbf{31.02}$ & 31.21 \\
{\bf Type II} & 27.26 & 27.83 & 28.65 & 31.01 & $\mathbf{31.22}$ \\
{\bf Type III} & 27.21 & 27.74 & 28.56 & 30.88 & 31.14 \\
\hline
\end{tabular}
\end{table}

We further investigate the impact of the quantization bit allocation strategy, discussed in {\bf Remark 1}, on the performance of the proposed frameworks. To this end, in Table~\ref{tab:psnr_bitbudget}, we compare the PSNR performance of the proposed MSVQ-SC framework under different bit allocation strategies, summarized as follows:
\begin{itemize}
    \item {\bf Type I}: The first 64 VQ modules, corresponding to sub-vectors with higher variances, are assigned 8, 7, and 6 bits for the first, second, and third stages, respectively. The remaining 64 VQ modules are assigned 6, 5, and 4 bits for the first, second, and third stages, respectively.

    \item {\bf Type II}: The first 64 VQ modules, corresponding to sub-vectors with higher variances, are assigned 7 bits across all stages, while the remaining 64 VQ modules are assigned 5 bits across all stages.  

    \item {\bf Type III}: All VQ modules are uniformly assigned 6 bits across all stages.  

\end{itemize}
The results in Table~\ref{tab:psnr_bitbudget} demonstrate that allocating more quantization bits to high-variance sub-vectors ({\bf Type~I} and {\bf Type~II}) yields performance improvements compared to uniform allocation ({\bf Type~III}). Furthermore, in the medium-to-low overhead regime, assigning more bits to earlier stages ({\bf Type~I}) provides better reconstruction performance than uniformly distributing bits across stages ({\bf Type~II}). These findings validate our variance-guided and stage-aware allocation principle in {\bf Remark 1}, confirming its effectiveness as a practical design guideline for the proposed frameworks.


\begin{figure}[t]
    \centering
    {\epsfig{file=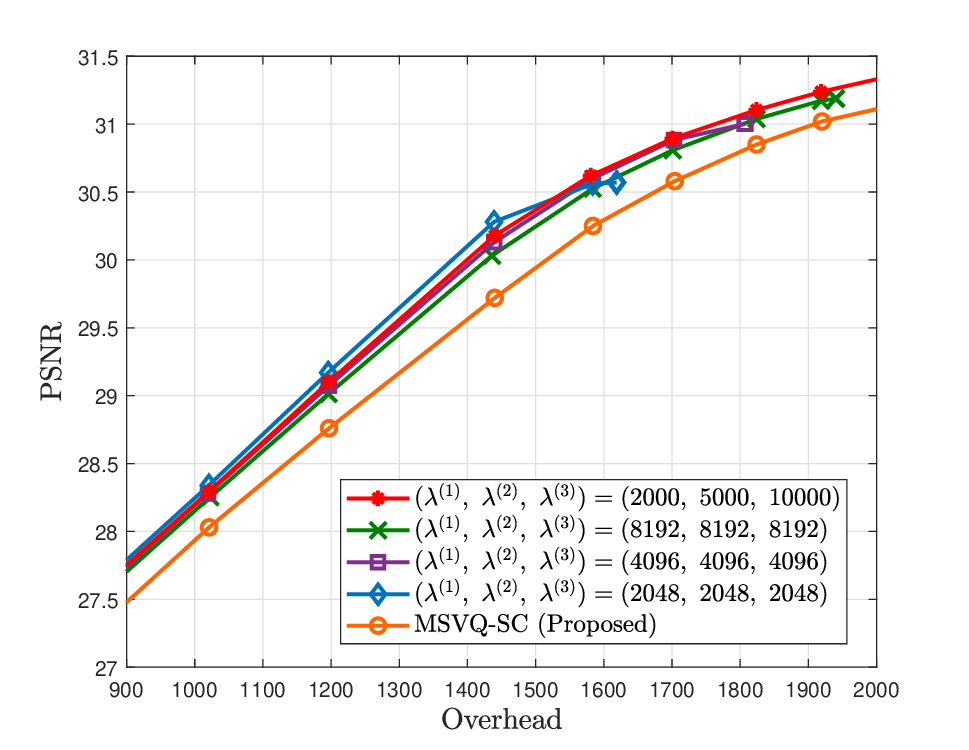,width=9cm}}
    \caption{Comparison of the PSNR performance of the proposed EC-MSVQ-SC framework with different values of (${\lambda}^{(1)}$, ${\lambda}^{(2)}$, ${\lambda}^{(3)}$).}\vspace{-3mm}
    \label{fig:PSNR_lmbdavsOverhead}
\end{figure}

We also investigate the effectiveness of the rate–distortion-aware modification introduced in Sec.~V-A for the proposed EC-MSVQ-SC framework. To this end, in Fig.~\ref{fig:PSNR_lmbdavsOverhead}, we compare the PSNR performance of the proposed EC-MSVQ-SC framework with different values of $\lambda^{(t)}$ in \eqref{eq:Modified quantize}. As a performance baseline, we also plot the PSNR of the proposed MSVQ-SC framework. Recall that a larger value of $\lambda^{(t)}$ places greater emphasis on the distortion term, causing the performance to approach that of the scheme without entropy coding. Conversely, a smaller $\lambda^{(t)}$ increases the relative weight of the rate term, reducing the influence of distortion. Although this leads to larger quantization errors, it significantly decreases the bit overhead. Fig.~\ref{fig:PSNR_lmbdavsOverhead} shows that when $\lambda^{(t)} = 8192$, the distortion term dominates, yielding reconstruction performance comparable to that of the framework without entropy coding, with negligible degradation, while still achieving a reduction in bit overhead. In contrast, for $\lambda^{(t)} = 4096$ and $\lambda^{(t)} = 2048$, the increased emphasis on the rate term results in greater reductions in bit overhead, but at the expense of higher quantization error, leading to more pronounced degradation in reconstruction performance in the high-overhead regime. These results highlight the inherent trade-off between rate efficiency and distortion, and demonstrate that the proposed framework can flexibly balance the two by adjusting the parameter $\lambda^{(t)}$.


\subsection{Evaluation under Realistic Communication Scenarios}\label{Communication performance}

\begin{figure}[t]
    \centering
    {\epsfig{file=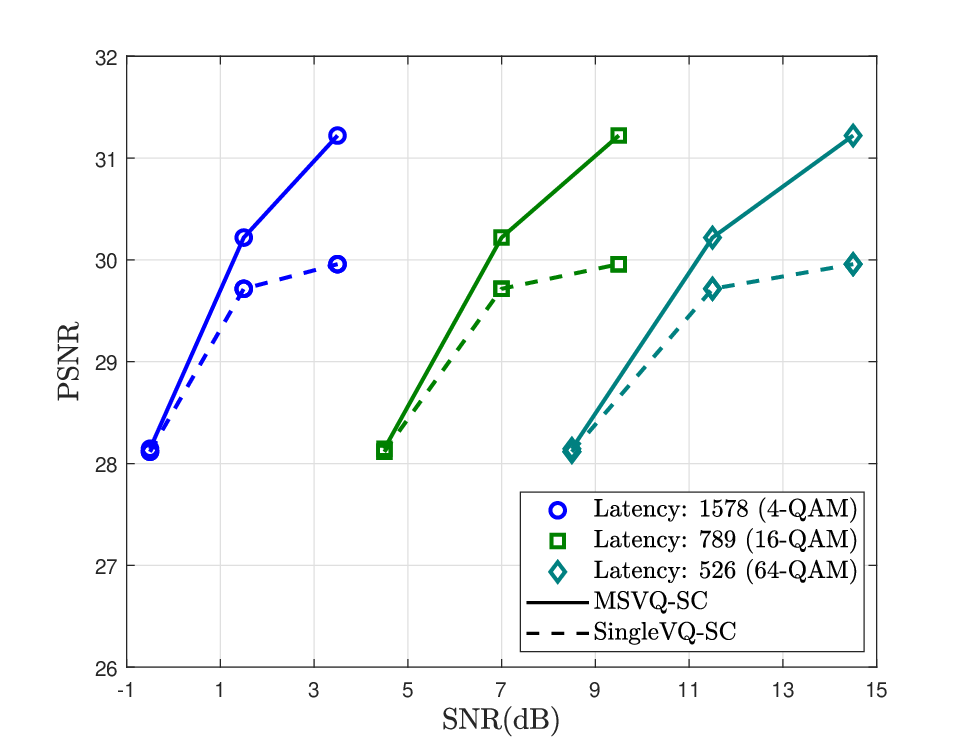,width=9cm}}
    \caption{Comparison of the PSNR performances of various digital SC frameworks with different latencies (symbols/image) under the realistic communication scenario.}\vspace{-3mm}
    \label{fig:rate_adaptation}
\end{figure}

In this subsection, we evaluate the superiority of the proposed frameworks under a realistic communication scenario in which standard channel coding and modulation schemes are employed. Specifically, we consider LDPC coding with block length $3156$ and three different code rates ($1/3$, $1/2$, and $2/3$), in combination with three modulation schemes: 4-QAM, 16-QAM, and 64-QAM. The latency of realistic communications is defined as the number of modulated symbols required to transmit a single image, measured in symbols per image.
For the proposed MSVQ-SC framework, the number of information bits is set to $1052$, $1578$, and $2104$ bits for code rates of $1/3$, $1/2$, and $2/3$, respectively. Unlike the proposed framework, the transmission rate of the existing SingleVQ-SC framework cannot be adjusted in a fine-grained manner. Therefore, we adjust its codebook size so that the number of information bits is approximately 1024, 1536, and 2048 bits for code rates of $1/3$, $1/2$, and $2/3$, respectively.
As shown in Fig.~\ref{fig:rate_adaptation}, the proposed framework consistently outperforms SingleVQ-SC while enabling precise rate adaptation across diverse communication scenarios. These results highlight that the proposed framework can be seamlessly integrated into standard wireless systems while retaining its rate-adaptive capability.




\section{Conclusions}\label{Sec:Conclusions}
This paper has introduced MSVQ-SC, a novel rate-adaptive SC framework that enables fine-grained rate control by selectively activating multi-stage VQ modules. Unlike conventional single-stage VQ methods, the proposed framework reduces computational complexity and mitigates codebook collapse by employing the MSVQ architecture. Meanwhile, the proposed framework also optimizes task performance under rate constraints via the stage selection optimization. Extensions with entropy coding and shared codebooks further improve transmission efficiency and reduce memory requirements. Simulation results have confirmed that the proposed framework consistently outperforms existing digital SC frameworks across multiple metrics, while offering robust adaptability to varying bit budgets and seamless integration with standard wireless systems. These findings establish MSVQ-SC as a practical and scalable solution for efficient rate-adaptive SC. 

Future research may extend the proposed MSVQ-SC framework in two directions. 
First, an adaptive grouping strategy can be developed where each group is assigned a different number of subvectors depending on their statistical characteristics or importance. 
Such non-uniform grouping is expected to improve codebook utilization and enhance overall transmission efficiency. Second, the inherent stage-wise design of the MSVQ-SC framework can be further exploited to realize a feedback-based retransmission scheme. By allowing the receiver to selectively request retransmission of specific stages, the framework could enhance reliability while preserving efficient resource usage.

\end{document}